%% file: odyssey2018_article_arxiv.tex
\documentclass[a4paper]{article}
\usepackage{Odyssey2018}
\usepackage{epsfig,amssymb,amsmath}
\ninept

\usepackage[]{algorithm2e, color}
\usepackage{multirow}
\usepackage{url}

\title{Weakly Supervised Training of Speaker Identification Models}

\name{Martin Karu, Tanel Alum\"{a}e}

\address{Department of Software Science  \\
Tallinn University of Technology, Estonia \\
{\small \tt martinkaru@gmail.com, tanel.alumae@ttu.ee} }
\begin{document}
\maketitle

\begin{abstract}
We propose an approach for training speaker identification models in a weakly supervised manner. We concentrate on the setting where the training data consists of a set of audio recordings and the speaker annotation is provided only at the recording level. The method uses speaker diarization to find unique speakers in each recording, and i-vectors to project the speech of each speaker to a fixed-dimensional vector. A neural network is then trained to map i-vectors to speakers, using a special objective function that allows to optimize the model using recording-level speaker labels. We report experiments on two different real-world datasets. On the VoxCeleb dataset, the method provides 94.6\% accuracy on a closed set speaker identification task, surpassing the baseline performance by a large margin. On an Estonian broadcast news dataset, the method provides 66\% time-weighted speaker identification recall at 93\% precision. 
\end{abstract}

\section{Introduction}

Conventional speaker identification models are usually trained on data where the speech segments corresponding to the target speakers are hand-annotated. However, the process of hand-labelling speech data is expensive and doesn't scale well, especially if a large set of speakers needs to be covered. This makes such models difficult to manage and to deploy. For example, this problem is evident in the domain of media monitoring or indexing of large speech corpora where one might need to identify the speech segments of a wide and growing range of public figures, such as politicians, scientists and celebrities.

On the other hand, very large collections of audio and video documents are available online. The documents often come with metadata which may provide information about the persons speaking in the document. For example, many items in the BBC radio archive\footnote{\url{http://www.bbc.co.uk/archive/}} have a `Contributors' section which lists the names of the speakers appearing in the programme, often together with names of other crew, such as producers and directors. Often, the names of the speakers appearing in the audio/video document are given in free form text. For example, a video on YouTube that contains an interview with Elon Musk has probably the name `Elon Musk' in the title and may further elaborate on the contents in the video description. However, the metadata rarely provides information about `who spoke when', i.e., it doesn't provide the necessary information for training speaker identification models in a supervised way.

This paper presents a method for training speaker identification models in a weakly supervised manner, relying only on the possibly incomplete set of speakers occurring in each audio document in the training set. That is, the method does not need training data annotation at the speech segment level but rather at the recording level. The method relies on speaker diarization, i-vectors \cite{dehak2011front} and deep neural networks. First, each recording in the training corpus is processed by a speaker diarization module that partitions the recording into homogeneous segments, applies speech/non-speech detection and clusters the resulting speech segments according to likely (anonymous) speaker turns. Second, i-vectors are extracted from each segment and averaged across the speaker turns. Finally, a deep neural network (DNN) is trained to perform speaker identification, using i-vectors as inputs and true speaker identities as outputs. 
Since we don't know the true mapping between i-vectors and speakers, we cannot use supervised learning using the cross-entropy criterion to train the DNN, as it is usually done when training classification models. Instead, we use a technique inspired by a method called expectation regularization \cite{mann2007}: rather than providing the true speaker label for each i-vector, we provide a speaker distribution over allowed speaker labels for the i-vectors of each recording and the training procedure is encouraged to find model parameters that predict a similar speaker distribution for each recording. That is, the loss function is defined at the recording level, not at the speaker level.
If different speakers occur intermixed in different recordings in the training data, then the optimal solution to this objective function is a model that predicts the true label for each i-vector. The method is surprisingly robust and easy to implement: it requires defining a loss function for the DNN that can be implemented in just a few lines of code in any modern deep learning framework.

The evaluation of this technique is performed on two very different datasets: the VoxCeleb dataset of YouTube videos \cite{nagrani2017} and an Estonian broadcast news dataset. VoxCeleb contains automatically retrieved videos corresponding to over 1000 different US celebrities, with around 18 videos per person. The videos do not have any metadata, other than the name of celebrity  that was used for the corresponding YouTube search. The Estonian broadcast news dataset consists of over 6000 recordings of the main evening news program of Estonian national radio. Each recording is accompanied with metadata that lists all speakers (both news reporters and interviewees) speaking in that programme. The two datasets are very different: while the VoxCeleb dataset is relatively heterogeneous, with unconstrained audio recording conditions, it contains roughly equal amount of speech for each speaker identity. The Estonian broadcast news database is much more homogeneous but the amount of speech from different speakers varies greatly: some news reporters appear in thousands of recordings while most speakers occur only once. We show that the proposed method works well for both datasets. 

The remainder of the paper is organized as follows. Section \ref{sec:related} presents related work. Section \ref{sec:method} describes the main idea of the training algorithm. Experimental results are shown in section \ref{sec:exp}. Section \ref{sec:conclusion} concludes the paper.

\section{Related work}
\label{sec:related}

\subsection{Named speaker identification from speech transcripts}

An alternative way for avoiding costly segment-level annotations for large-scale speaker name identification in audio documents is to use automatically generated speech transcripts to find hints about the speaker identities. This principle is based on the assumption that the names of the speakers are pronounced at some point in the document, as it usually happens in broadcast news. The general principle behind this idea is to detect named entities in speech transcripts and determine whether they refer to a speaker in the document or not. This decision is usually done based on linguistic cues. A popular approach is based on semantic classification trees (SCT) \cite{mauclair2006, elkhoury2012}  that are trained to automatically learn lexical rules that determine whether a  detected named entity is associated with the current speaker turn, previous turn, next turn or neither of them. (e.g. ``thank you Christiane Amanpour'' probably means that the name corresponds to the previous speaker).
The rules are learned based on a large manually transcribed training corpus where speaker names are annotated for each speaker turn. The method was tested on the  ESTER corpus of French broadcast news and was shown to be able to identify the correct speaker during 70\% of the total speech duration (with 18\% of false identifications) when manual speaker diarization and manual speech transcripts are used \cite{mauclair2006}. Later it was shown that using automatic speaker diarization and automatic transcripts has a big negative effect on the accuracy of transcript based named identification, increasing the overall duration of false identification from 17\% to 75\% \cite{jousse2009}. 

One of the drawbacks of this approach is that the linguistic patterns of using names for speaker introduction (or self-introduction) are highly domain-specific. For example, in broadcast news, reporters and interviewees are introduced using different words than in talkshows. 

\subsection{Named speaker identification from TV captions}

When working with video data, written names in the overlaid video captions can be used as another source of information for named speaker identification. Early experiments with this approach struggled with high name detection error rate due to the bad video quality \cite{satoh1999}. Later approaches identify names with very high precision, partly thanks to better video quality and partly due to improved algorithms. For example, in \cite{poignant2015}, 80\% precision  and 68\% recall was reported on the REPERE corpus of French TV broadcasts, using time-weighted error metrics.

The main shortcoming of this approach is that it only works for broadcast TV data. It cannot be applied to radio broadcasts and to online video archives where speaker names are not annotated in the overlaid title block.

\subsection{Named speaker identification in video using face identification}
\label{sec:voxceleb}

When working with video data, speakers could also be identified based on face identification. This approach was used for constructing the VoxCeleb database \cite{nagrani2017} which contains hundreds of thousands of 'real
world' utterances for over 1000 celebrities. The database was collected from YouTube, using the following workflow. First, a target speaker list was constructed, using an intersection of popular celebrity names and persons appearing in the VGG Face dataset \cite{parkhi2015}. Second, 50 top YouTube videos were retrieved for each person, using a query ``$<$name$>$ interview''. Third, a face detector was used to identify faces in video frames. Fourth, audio-video synchronization between detected faces and speech was determined, in order to identify which face in the video corresponds to the current speaker (if any). Finally, a face verification system was used to determine whether the face of the active speaker corresponds to the target person of the video. This results in a database where high-confidence video segments corresponding to the target speakers are automatically annotated and can be used for training speaker identification models. Experiments showed that a speaker identification model trained on such data achieves 80.5\% accuracy, using a closed set identification task.

\subsection{Multi-instance multi-label learning}

The task of training a speaker identification model based on speakers annotated at the recording level could be formulated as a multi-instance multi-label (MIML) learning problem \cite{zhou2007multi}. MIML handles the general classification task where objects in the training set are decomposed into a bag of instances and are associated with multiple class labels. For example, in image classification problems, an image could be decomposed into segments and labeled with multiple labels, whereas the relationship between segments and labels is not explicitly given in training data. The standard goal in MIML is to learn a classifier that predicts a label set for an unseen bag of instances. Various algorithms for MIML have been proposed in recent years \cite{zhou2007multi, zhou2012multi, zhang2014review}. 

A common approach to MIML shared by several recently proposed algorithms \cite{huang2014fast, briggs2012rank} is to use a ranking loss to optimize the MIML classifier. For example, the MIMLfast algorithm \cite{huang2014fast} uses the following method to learn the model: at each step, MIMLfast randomly samples a triplet from training data which consists of a bag, a relevant label of the bag and an irrelevant label, and optimizes the model to rank the relevant label before the irrelevant one if such an order is violated.

The method proposed in this paper could be extended to handle the general MIML problem and is one of the directions for our future work.

\subsection{Label regularization}
\label{sec:expreg}

Our method is directly inspired by a technique called expectation regularization \cite{mann2007}, a semi-supervised method that augments the traditional log-likelihood loss function with an additional term that encourages the model predictions on unsupervised data to match some expectations, typically given as class priors. The idea of this method is that while the mapping between inputs and outputs must be learned from supervised data, prior knowledge about the unsupervised data can provide the model important side information. The method penalizes models by divergence between the model's average expectations over the unsupervised data and prior probabilities of the labels, which can be estimated from labeled data or given as prior knowledge. The method is found to be robust, it is easy to implement and scales well to large datasets.

\section{Proposed method}
\label{sec:method}

The proposed method relies on an annotated set of speakers appearing in each audio recording. Note that the set does not need to be exhaustive: only the speakers that need to be identifiable by the system must be included in the sets. 

\begin{figure}[tb]
	\includegraphics[width=\columnwidth]{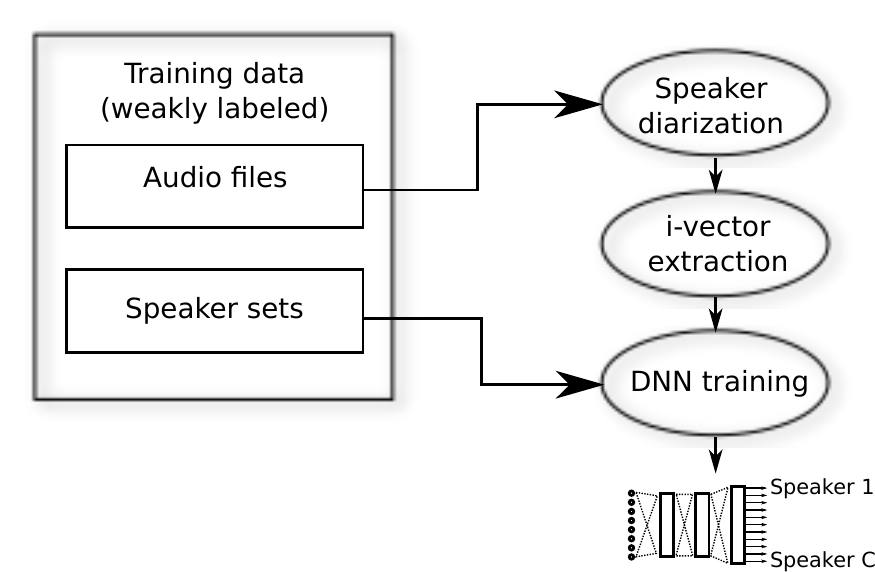}
	\caption{{\it Architecture of the weakly-supervised training process.}}
	\label{fig:architecture}
\end{figure}

Outline of the training process is depicted in Figure \ref{fig:architecture}.
First, we apply a speaker diarization system to the training data. This step partitions the recording into homogeneous segments, discards non-speech segments and clusters the speech segments that are likely uttered by the same speaker. 

Next, we use an i-vector extractor to compute i-vectors for all speakers in all recordings. The i-vector extractor can be trained on available labeled training data, possibly from another domain. Alternatively, the i-vector extractor can be trained on the automatically clustered speakers of the training set. We experimented with both methods and found no clear difference between those alternatives.

Here we give some useful notations for the rest of this section. Let $\mathcal{X} \subset R^D$ denote the $D$-dimensional feature space (i.e., the i-vector space) and $\mathcal{Y} = \{1, 2, ..., C\}$ the set of target speaker identities. The training corpus $\mathcal{D}$ contains a set of $N$ audio recordings $\{X_n\}_{n=1,2,...,N}$ where each recording $X_n$ contains a set of i-vectors for the diarized speakers $X_n=\{x_{mn}\}_{m=1,2,...,M_n}$, with $x_{mn} \in \mathcal{X}$. The training corpus also contains the corresponding sets of speaker labels for each recording $\{Y_n\}_{n=1,2,...,N}$ where $Y_n \subset \mathcal{Y}$. Note that the number of speaker labels does not have to agree with number of i-vectors for the corresponding recording, and the correspondence between  $x_{mn}$ and the elements in $\{Y_n\}$ is not known.
The task is to train a model to classify i-vectors based on the speaker identities, i.e., to learn a classification function $f: \mathcal{X} \rightarrow \mathcal{Y}$ from the training dataset.

We use a feed-forward neural network to learn the mapping from the i-vector space to the speaker label space. First, we augment the set of target speaker identities with a special $\langle unk \rangle$ identity reserved for unknown speakers, $\mathcal{Y}' = \mathcal{Y} \cup \{\langle unk \rangle\}$. The neural network takes i-vectors as input and has $|\mathcal{Y}'|$ outputs. The softmax function is used in the final layer of a neural network.

The neural network is trained using an objective function defined at the recording level, not at the single training sample level as usually. Specifically, we want the neural network predict a similar set of speakers for the set of i-vectors as is defined in the metadata. To achieve this, we first define the expected average distribution over the speaker labels for each recording as 
\begin{equation}
\bar{\tilde{p}}_n(y_i) = \left\{
  \begin{array}{@{}ll@{}}
    \frac{1}{|X_n|}, & \text{if}\ y_i \in Y_n \\
    max(0, 1-\frac{|Y_n|}{|X_n|}), & \text{if}\ y_i = \langle unk \rangle\  \\
    0, & \text{otherwise}
  \end{array}\right.
\label{eq:exp}
\end{equation}

The recording level objective function is the Kullback-Leibler divergence between the expected average distribution and model's expected average conditional distribution:
\begin{equation}
D(\bar{\tilde{p}}||\bar{\hat{p}}_{\theta}) = \sum_y \bar{\tilde{p}} \log \frac{\bar{\tilde{p}}}{\bar{\hat{p}}_{\theta}} 
\end{equation}

The idea of this objective function is that we don't know the exact correspondence between the i-vectors and speakers of a recording, but we want exactly a single i-vector to be assigned to each speaker of the show, and the rest (if any) to be absorbed by the class that is reserved for unknown speakers.  Clearly, the assumption that only one i-vector corresponds to a labeled speaker is not always true: in broadcast news, a news anchor could be speaking at the beginning of the news show with background music, and later without music, which usually causes the speaker diarization module to split the single speaker into two pseudo-speakers. Similarly, a reporter could speak both in a studio setting and with a noisy background in a single news show, also resulting in two i-vectors. However, we haven't found this to cause any noticeable problems.

The proposed method works only if the recordings in the training data have sufficiently different speaker distributions. Note that the objective function does not encourage the neural network in any way to assign a high probability to a particular speaker in the recording -- given a single recording and a set of speakers for that recording, the objective function can be minimized by predicting a uniform distribution over all i-vectors for all speakers appearing in that particular recording. However, since we require the recordings to have different speaker distributions, the neural network has to start assigning non-uniform distributions to the i-vectors of a show, in order to better fit the data.

The neural network also doesn't learn to differentiate between the speakers that occur always together in the same recordings. 

The training algorithm is summarized in listing \ref{alg:training}.

\begin{figure}[t]
	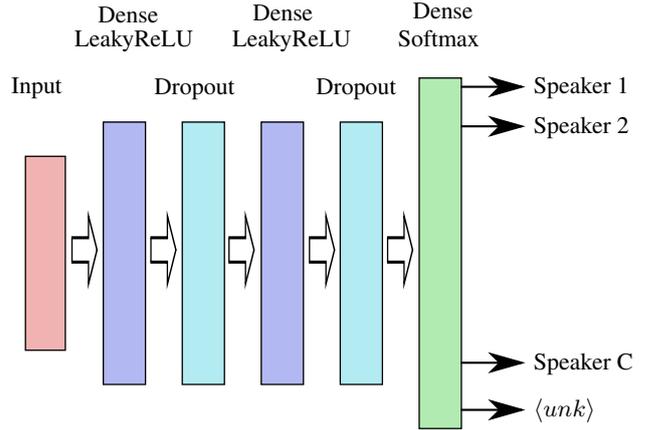
	\caption{{\it Architecture of the speaker identification DNN.}}
	\label{fig:dnn}
\end{figure}

We used a very simple deep neural network as the underlying model (Figure \ref{fig:dnn}) . The network has two fully-connected hidden layers, using the leaky rectified linear units. The number of hidden units was optimized on development data. We found that dropout layers added after the dense layers improve performance of the model. 

\begin{algorithm}[tp]
	\KwData{\\
	List of target speaker names: $\mathcal{Y} = \{1, 2, ..., C\}$ \;
	A set of $N$ diarized audio recordings with metadata, consisting of 
	\begin{itemize}
		\item 	i-vectors $\{X_1, ..., X_N\}$, $X_n = \{x_{mn}\}$ 
		\item 	Speaker names: $\{Y_1, ..., Y_N\}, Y_n \subset \mathcal{Y}$ 
	\end{itemize}
    } 
	\KwResult{Trained model that maps $\mathcal{X} \rightarrow \mathcal{Y}$}
	Initialize neural network with parameters $\theta$\;
	Precompute expected average speaker distribution $\bar{\tilde{p}}_n$ for each recording according to eq. \ref{eq:exp}\;
	\While{training hasn't converged}{
		Shuffle training data\;
		\For{$n = 1...N$}{
			Compute the neural network predictions $\hat{p}_{\theta}^m$ for i-vectors $x_{mn}$\;
			Compute average predictions $\bar{\hat{p}}_{\theta} = \frac{1}{M}\sum_{m=1..M} \hat{p}_{\theta}^m$\;
			Update model weights: $\theta \leftarrow SGD(D(\bar{\tilde{p}}_n||\bar{\hat{p}}_{\theta}))$\;
		}
}
\caption{Algorithm for training a neural network on weakly labeled data.}
\label{alg:training}
\end{algorithm}

This method is very similar to expectation regularization, described in section \ref{sec:expreg}. However, we use it without the supervised log-likelihood loss, and we use a different prior probability for each minibatch (where the minibatches correspond to the i-vectors of the same recording), not globally as in expectation regularization.

\section{Experiments}
\label{sec:exp}

We conducted experiments on two datasets: the VoxCeleb database and an Estonian broadcast news dataset.

\subsection{VoxCeleb}

As described in section \ref{sec:voxceleb}, VoxCeleb is a dataset that contains YouTube videos corresponding to over 1000 celebrities. The dataset is collected automatically by retrieving top matches from YouTube for queries ``$<$name$>$ interview'', for a predefined list of celebrity names.

\begin{table}[tbh]
	\caption{{\it Statistics of the VoxCeleb dataset. Medians shown with lower and upper quartiles where appropriate.}}
	\label{tbl:voxceleb_stat}
	\vspace{2mm}
	\centerline{
		\begin{tabular}{|p{5cm}|c|}
			\hline
			\textbf{\# of target speakers} & 1251 \\
			\textbf{\# of videos per target speaker}  & 14 (13 / 22) \\
			\textbf{Video length (s)}  & 268 (171 / 426)\\
			\textbf{\# of diarized speakers per video }& 3 (2 / 4) \\
			\hline
		\end{tabular}}
	\end{table}

One method to train a speaker identification system based on this dataset was proposed in \cite{nagrani2017} and summarized in section  \ref{sec:voxceleb}. The method relies on a large face recognition database that is used to train a face verification model that covers all the celebrities in the dataset. The face verification model is used to identify video segments where the corresponding celebrity is speaking and the speaker identification model is trained based on those speech segments. 

Our method, on the other hand, uses only the speech data. Additionally, the name of the celebrity   for whom the video corresponds to is used as weak supervision\footnote{We found that there were some videos that were retrieved for several celebrities. In this case, we used all the names as weak supervision for this video}.

We used a speaker diarization system trained on Estonian radio broadcast data \cite{alumae2014recent} to segment and cluster the VoxCeleb training data, despite the obvious domain mismatch. The diarization system is based on the LIUM SpkDiarization toolkit \cite{meignier2010lium} and uses BIC clustering \cite{chen1998} followed by CLR-like clustering \cite{reynolds1998blind} to find the most likely segment-to-speaker mappings. 

The speaker-diarized VoxCeleb training data was used to train an i-vector extractor using Kaldi \cite{kaldi}. We use 600-dimensional i-vectors. The underlying 20-dimensional MFCC features are extracted from band-limited speech signal, with the high cutoff frequency of 3700 and a frame length of 20. Speaker-level i-vectors are computed by length-normalizing the utterance i-vectors, averaging the utterance vectors and finally length-normalizing the average. 

The DNN for speaker identification has two hidden layers with a dimensionality of 1024. Dropout layers use a dropout proportion of 0.2 at training time. The DNN was trained for 50 epochs using the described weakly supervised method, with a linearly decreasing learning rate on the training partition of the VoxCeleb data. 

Validation and evaluation is performed on the segments of the VoxCeleb development and test data that are verified to belong to the corresponding celebrity using face identification. That is, we use the same speech segments for speaker identification evaluation as are used in the VoxCeleb paper. 

Although our speaker identification model is an open-set model, meaning that it assigns a probability to the prediction that the input is an an unknown speaker, we use the model at test time in a closed set mode by discarding the unknown speaker output. This is compatible to the approach used in the VoxCeleb paper.

The speaker identification results are given in Table \ref{tbl:voxceleb}. Our DNN trained in weakly-supervised manner achieves 94.6\% \textit{top-1} accuracy, noticeably higher than the CNN trained on face-verified segments, as proposed in the VoxCeleb paper.

\begin{table}[th]
	\caption{\label{table1} {\it Speaker identification accuracies on the VoxCeleb test set.}}
    \label{tbl:voxceleb}
	\vspace{2mm}
	\centerline{
		\begin{tabular}{|p{4cm}|c|c|}
			\hline
			\textbf{System} &\textbf{ Top-1 (\%)} &\textbf{ Top-5 (\%)} \\
			\hline  \hline
			CNN on spectrograms, trained on face-verified segments \cite{nagrani2017} & 80.5 & 92.1 \\
			DNN on i-vectors, weakly supervised   & 94.6 & 98.1 \\
			\hline
		\end{tabular}}
	\end{table}

\subsection{Estonian broadcast news}

In the second experiment, we use recordings of the Estonian Public Broadcasting main evening news radio programme, \textit{P\"{a}evakaja}. The programme delivers daily news reports and commentary from people involved.

\subsubsection{Training data}
In order to perform weakly supervised training, we downloaded all available 6619 recordings of the programme from the archive of the broadcasting organization. The programme has been on air since 1944 but the majority (92\%) of the recordings in the archive originate from 2004 and later. As we executed the download at the end of 2016, the last show originates from November 16, 2016.
 Most of
the recordings range from 10 to 24 minutes in length. A small percentage of recordings is less than 5 minutes or more than 25 minutes long. The total duration of the recordings is 1854 hours.

\begin{figure}[tb]
\includegraphics[width=\columnwidth]{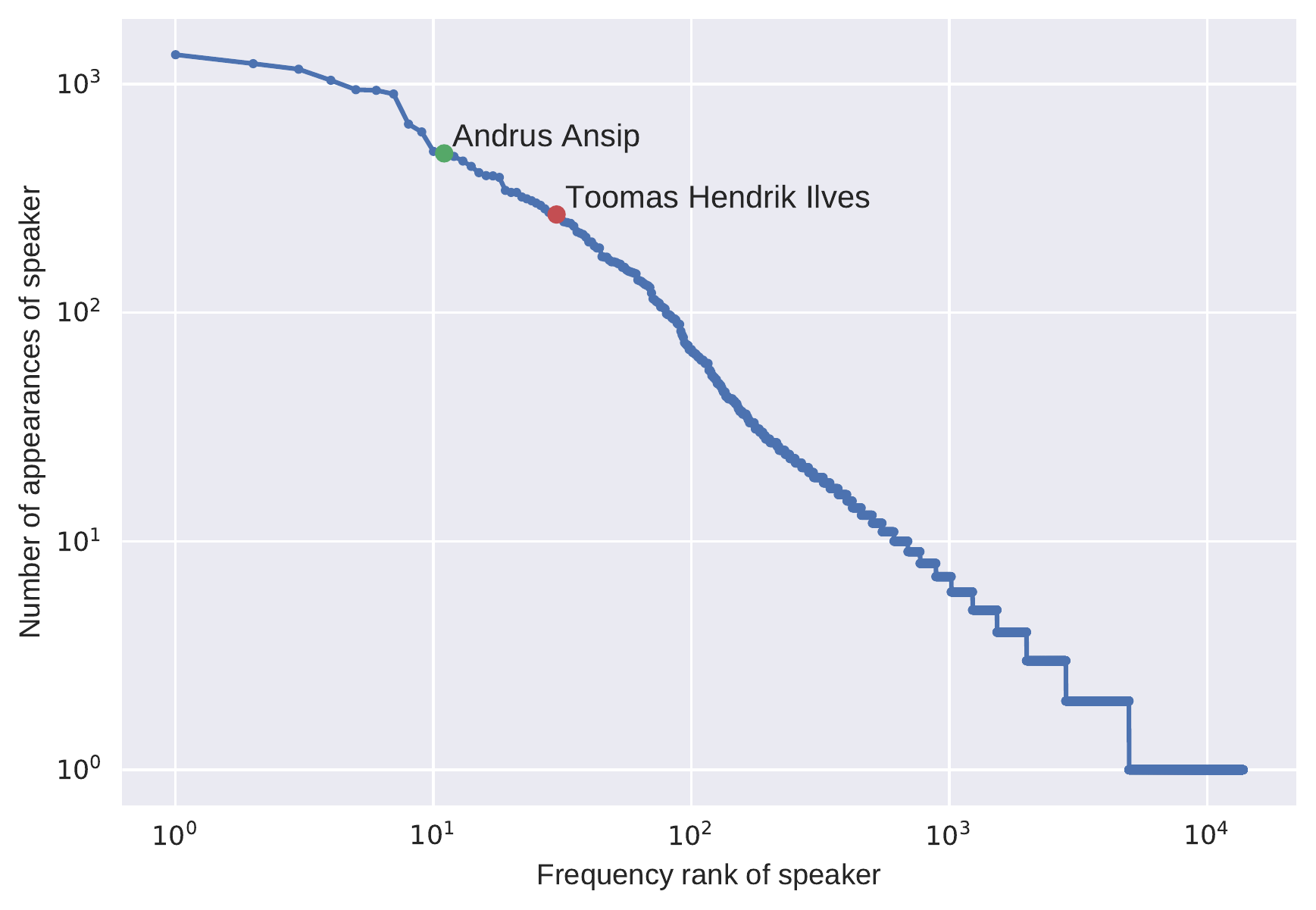}
\caption{{\it Zipf plot showing the number of appearances per speaker in  \textit{P\"{a}evakaja} training data \textit{vs}. his/her frequency rank.}}
\label{fig:zipf}
\end{figure}

The recordings in the archive are accompanied with metadata that also includes the names of the speakers that appear in the programme. The number of different names across all programmes is 13\,771. The majority of speakers occur only once: only 4967 appear more than once, and 1529 appear at least five times. The most frequently appearing speakers are news reporters and anchors. The most frequent speaker who is not related to the production of the programme is Andrus Ansip  who was the prime minister of Estonia from 2005 to 2014. He appears in 498 different shows. Figure \ref{fig:zipf} shows the number of appearances \textit{vs} the ranks of all speakers in the training data. The positions of Andrus Ansip and Toomas Hendrik Ilves (the president of Estonia from 2006 to 2016) are highlighed.

\begin{figure}[tb]
\includegraphics[width=\columnwidth]{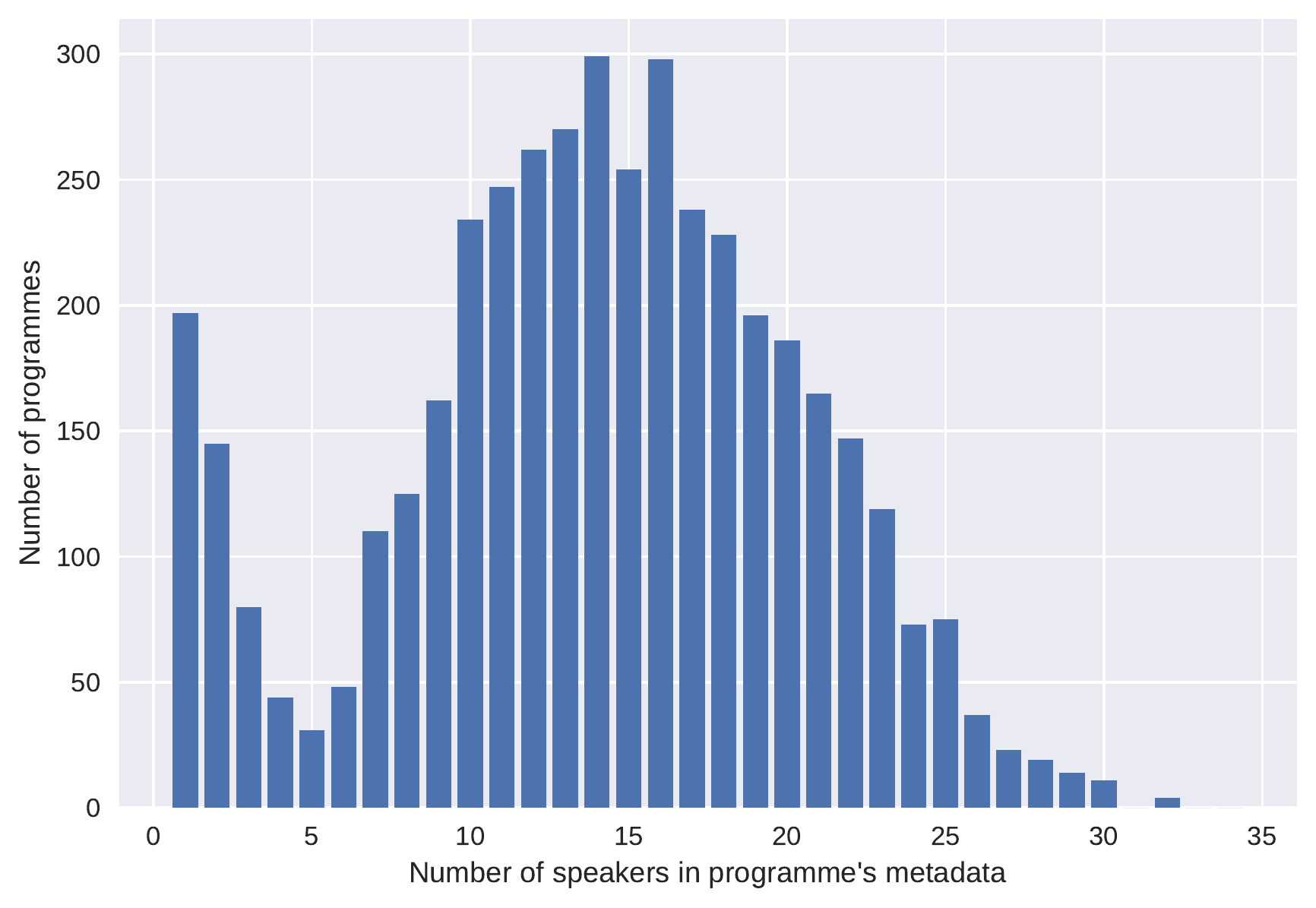}
\caption{{\it Histogram of the number of speaker per show's metadata in  \textit{P\"{a}evakaja} training data, after discarding shows with no speaker annotation.}}
\label{fig:histogram}
\end{figure}

When analyzing the speaker information given in the metadata, we discovered that the metadata of over 2000 different shows not include any speaker information. It turned out that most of the shows between 2006 and 2008 miss the speaker annotation. We discarded the shows with zero speakers from our training data. The histogram of the number speakers per show (after discarding invalid shows) is shown in Figure \ref{fig:histogram}. We can observe two local maxima in the histogram: one corresponding to one speaker per show, and another one around the 15 speakers-per-show point. This can be explained by the fact that for some time periods, the archive also contains short news shows that often do consist of only a single news reporter speaking.

\subsubsection{Evaluation data}

For evaluation, we used 16 manually annotated news recordings of the same programme. The shows originate from June to July of 2017, that is, more than a half a year later than the last show in the training set. The first eight shows were used as a development set and the last eight shows for evaluation. Overview of the data is given in Table \ref{tbl:paevakaja_eval_stat}. The speakers appearing in the development and evaluation data were grouped into news anchors (including reporters and commentators) and other persons not related to the production of the programme. It can be seen that although only about one quarter of the speakers are anchors, their speech takes up the majority of the total speaking time.

\begin{table}[th]
\caption{\label{table1} {\it Development and evaluation set statistics.}}
\label{tbl:paevakaja_eval_stat}
\vspace{2mm}
\centerline{
\begin{tabular}{|p{5cm}|c|c|}
\hline
\textbf{} &\textbf{ Dev} &\textbf{ Eval} \\
\hline  \hline
\# shows & 8 & 8 \\
Total number of speakers & 99 & 121 \\
 \quad  Anchors & 24 & 32 \\
 \quad  Non-anchors & 75 & 89 \\
Total speaking time & 2h08 & 2h38 \\
 \quad  Anchors & 1h22 & 1h41 \\
 \quad  Non-anchors & 0h43 & 0h57 \\
Average show duration (min:sec) & 17:11 & 21:00 \\
Average \# of speakers per show & 16.1 & 19.0 \\ 
Average speaking time (seconds) & 60 & 62  \\
Minimum speaking time (seconds) & 6 & 8  \\
Maximum speaking time (seconds) & 320 & 401  \\
\hline
\end{tabular}}
\end{table}

\subsubsection{Data pre-processing}

For segmenting and clustering the training and evaluation data we used same speaker diarization system that was used in the VoxCeleb experiments. The mean diarization error rate (DER) of the system is 0.10 on development data and 0.12 on test data.
The i-vector extractor is trained on around 200 hours of manually annotated Estonian speech recognition training data. As in the VoxCeleb experiment, we use 600-dimensional i-vectors with length normalization.

\subsubsection{Model}

For training the speaker identification DNN, we prune the list of speakers to include only those that occur at least twice in the training data, resulting in 4939 unique names. This set covers 59\% of the development set speakers. The DNN has 1024 units in both hidden layers and uses a dropout proportion of 0.5. The model is trained for 100 epochs, with learning rate decreasing linearly from 0.01 to 0.001. The described hyperparameters were tuned so that the speaker recall of the development set at 95\% precision would be optimized. We optimized the model using the 95\% precision constraint as we intend to use it in an application where high precision is more important than high recall.

Note that contrary to the VoxCeleb experiment, the speaker identification in this task is an open set model, i.e., it assigns a probability to the unknown speaker.

\subsubsection{Results}

We used a speaker identification system trained in supervised manner for comparison. The system is trained on around 200 hours of manually transcribed speech data. The data consists of broadcast news, broadcast conversations (such as talk shows and interviews from both TV and radio), lectures and conference talks. Most of the material originates from between 2010 and 2014. The system includes models for 814 speakers -- all persons with at least 10 utterances in the training data. Cosine similarity metrics between speaker i-vectors is used for identifying speakers. The similarity threshold was optimized on development set so that recall at the 95\% precision would be optimized. Note that this actually results in 100\% precision on both development and test set, as any lower threshold would have reduced the precision below 95\%.

\begin{table}[tb]
	\caption{\label{table1} {\it Speaker identification results on manually segmented \textit{P\"{a}evakaja} data.}}
	\label{tbl:paevakaja_results1}
	\vspace{2mm}
	\centerline{
		\begin{tabular}{|l|c|c|c|c|}
			\hline
			\textbf{System} & \multicolumn{2}{c|}{\textbf{Dev}} & \multicolumn{2}{c|}{\textbf{Eval}} \\
			\hline
			& Precision & Recall & Precision & Recall \\
			\hline  \hline
			Supervised  & 100\% & 15\% & 100\% & 17\% \\
			Proposed & 95\% & 41\% & 95\% & 45\% \\
			\hline
		\end{tabular}}
	\end{table}

Table \ref{tbl:paevakaja_results1} lists speaker identification precision and recall of the baseline and proposed systems, using manually segmented evaluation data. The weakly supervised system greatly outperforms the baseline system. Actually, the proposed system even results in higher recall than the oracle maximum of the supervised system, since the supervised system covers only 28\% of the speakers in the development set.

It is difficult to compute speaker-level precision and recall metrics for the automatically segmented and diarized evaluation data, as a single real speaker can correspond to many clusters in the speaker diarization result, or vice versa. Therefore, we used time-weighted error metrics: identification error rate (IER), precision and recall characterize now the proportion of time during which the current speaker is identified correctly. We used the \textit{pyannote.metrics} Python package \cite{pyannote.metrics} to calculate the results. A collar of 0.5 seconds was used at speaker change boundaries. Time-weighted results on the evaluation set, based on both perfect (oracle) and automatic diarization, are listed in Table \ref{tbl:paevakaja_results2}. 

\begin{table}[tbh]
	\caption{\label{table1} {\it Time-weighted speaker identification results on manually segmented and automatically segmented \textit{P\"{a}evakaja} evaluation set.}}
	\label{tbl:paevakaja_results2}
	\vspace{2mm}
	\centerline{
		\begin{tabular}{|l|c|c|c|c|}
			\hline
			\textbf{Speakers} & \textbf{System} & \textbf{IER} & \textbf{Precision} & \textbf{Recall} \\
			\hline
			\hline
			\multicolumn{5}{|l|}{\textit{Oracle diarization}} \\
			\hline
			\multirow{ 2}{*}{All}  & Supervised & 49\% & 100\% & 51\% \\
			& Proposed & 28\% & 96\% & 75\% \\
			\hline
			\multirow{ 2}{*}{Anchors}  & Supervised & 22\% & 100\% & 78\% \\
			& Proposed & 4\% & 98\% & 98\% \\
			\hline
			\multirow{ 2}{*}{Non-anchors} & Supervised & 99\% & 100\% & 1\% \\
			& Proposed & 78\% &  89\% & 24\% \\
			\hline
			\hline
			\multicolumn{5}{|l|}{\textit{Automatic diarization}} \\
			\hline
			\multirow{ 2}{*}{All}  & Supervised & 56\% & 96\% & 45\% \\
			& Proposed & 35\% & 93\% & 66\% \\
			\hline
			\multirow{ 2}{*}{Anchors}  & Supervised & 34\% & 96\% & 69\% \\
			& Proposed & 14\% & 94\% & 89\% \\
			\hline
			\multirow{ 2}{*}{Non-anchors} & Supervised & 99\% & 100\% & 1\% \\
			& Proposed & 80\% &  90\% & 22\% \\
			\hline
		\end{tabular}}
	\end{table}

The proposed weakly supervised system performs notably better than the baseline supervised system that was used for comparison. The supervised system is not able to identify almost any of the non-anchor speakers, whereas the weakly supervised system identifies non-anchors correctly during 20\% of the time. This can be partly explained by the fact that most of the manually annotated training data originates from before 2015, and there has been a strong shift in Estonian political landscape since then, resulting in a different set of public figures who appear in radio news. This also shows the strength of the weakly supervised system, as it is much easier to obtain or create updated training data for it, compared to manually segmented and annotated data. 

Automatic speaker diarization errors have a noticeable effect on the accuracy of the speaker identification systems: the average identification error rate of the weakly supervised system increases from 28\% to 35\%. 

\begin{figure}[b]
\includegraphics[width=\columnwidth]{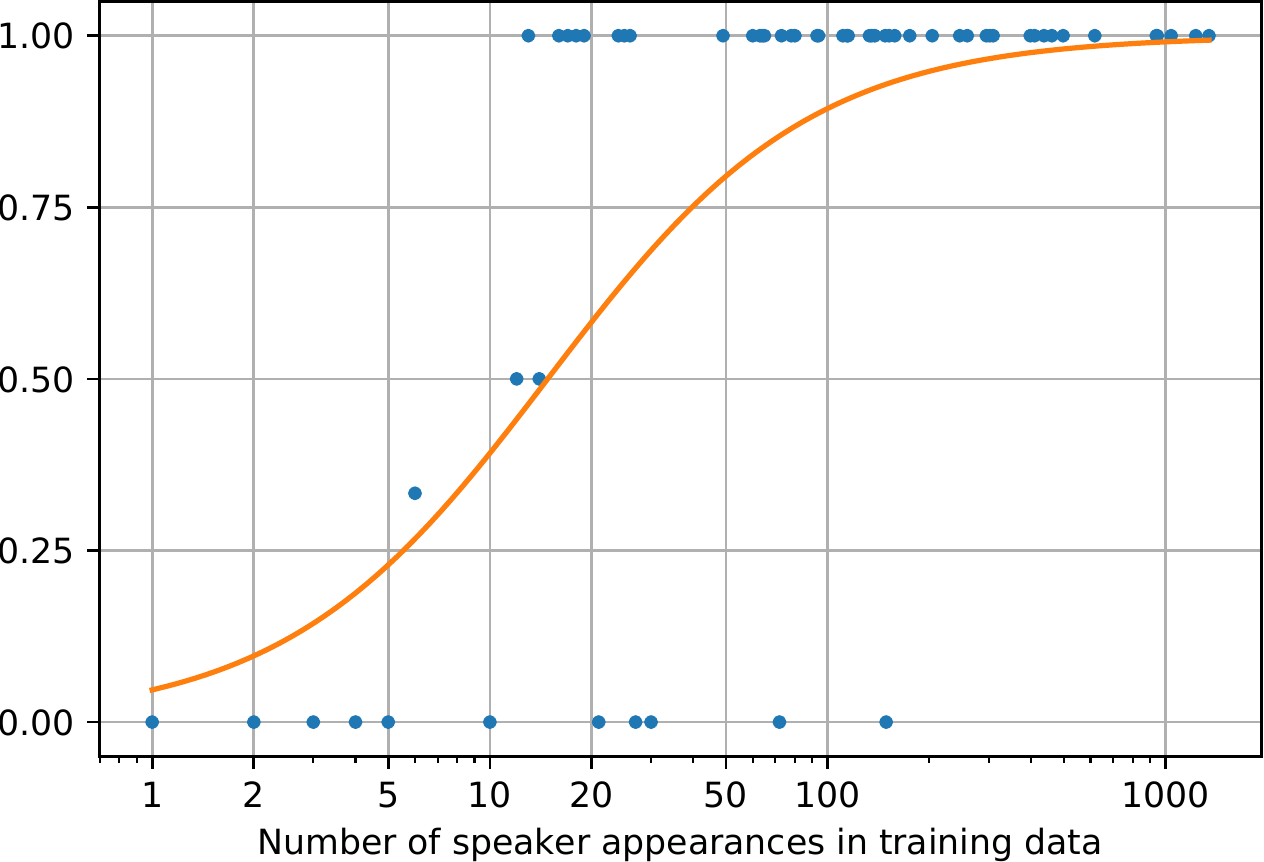}
\caption{{\it Number of appearances per speaker in training data \textit{vs} the probability of identifying him/her in evaluation data. The orange line shows a log-space fitted logistic regression curve.}}
\label{fig:curve}
\end{figure}

We analyzed how many times a speaker must appear in the weakly labeled training data in order to be identified in unseen data. Figure \ref{fig:curve} shows how the number of appearances of a speaker relates to the probability of him/her being correctly identified in unseen data. The dots correspond to the empirical identification probabilities (number of appearances \textit{vs} relative frequency of identification) and the line corresponds to the logistic regression curve fitted to the empirical data in log space. The dots show that, for example, speakers with 5 appearances in training data were never recognized in test data. The minimal number of appearances that resulted in correct identification was only six. However, around 15 appearances are needed for the probability of identification to become more than 50\%.

A speaker identification system based on the proposed approach is currently used in the  archive of automatically transcribed Estonian broadcast data \cite{tsab}, available at \url{http://bark.phon.ioc.ee/tsab}.

\section{Conclusion}
\label{sec:conclusion}

We have presented a method to train speaker identification models using only the information about speakers appearing in each of the recordings in training data, without any segment level annotation. Obtaining or creating such training data is much easier than segment-annotated data. To our knowledge, this is the first attempt to use such high-level annotation for this task.
During training, we use speaker diarization and i-vector extraction to map different speakers in each recording to fixed-dimensional vectors. A neural network is then trained using a special objective function that encourages the model to assign a similar average distribution of labels to the i-vectors of a single show as the annotation in the training data. Our experimental results show that a model trained using such weak labels can provide a recall of 66\% at  93\% precision on a broadcast news task. On the VoxCeleb dataset of YouTube videos, the method results in 94.6\% accuracy, greatly outperforming a baseline system that uses face identification for obtaining training data annotation.

In the future, we plan to extend the method to a more general multi-instance multi-label learning problem.

%\bibliographystyle{IEEEbib}
%\bibliography{odyssey2018_article.bib}

\input{odyssey2018_article_arxiv.bbl}
\end{document}

%% file: dnn.pdf_tex
%% Creator: Inkscape inkscape 0.91, www.inkscape.org
%% Accompanies image file 'dnn.pdf' (pdf, eps, ps)
\begingroup%
  \makeatletter%
  \providecommand\color[2][]{%
    \errmessage{(Inkscape) Color is used for the text in Inkscape, but the package 'color.sty' is not loaded}%
    \renewcommand\color[2][]{}%
  }%
  \providecommand\transparent[1]{%
    \errmessage{(Inkscape) Transparency is used (non-zero) for the text in Inkscape, but the package 'transparent.sty' is not loaded}%
    \renewcommand\transparent[1]{}%
  }%
  \providecommand\rotatebox[2]{#2}%
  \ifx\svgwidth\undefined%
    \setlength{\unitlength}{281.51339073bp}%
    \ifx\svgscale\undefined%
      \relax%
    \else%
      \setlength{\unitlength}{\unitlength * \real{\svgscale}}%
    \fi%
  \else%
    \setlength{\unitlength}{\svgwidth}%
  \fi%
  \global\let\svgwidth\undefined%
  \global\let\svgscale\undefined%
  \makeatother%
  \begin{picture}(1,0.56918091)%
    \put(0,0){\includegraphics[width=\unitlength,page=1]{dnn.pdf}}%
    \put(-0.00278906,0.44900177){\color[rgb]{0,0,0}\makebox(0,0)[lb]{\smash{Input}}}%
    \put(0.15207333,0.54338957){\color[rgb]{0,0,0}\makebox(0,0)[b]{\smash{Dense\\ }}}%
    \put(0.40115008,0.44697193){\color[rgb]{0,0,0}\makebox(0,0)[lb]{\smash{Dropout}}}%
    \put(0.3570877,0.54541942){\color[rgb]{0,0,0}\makebox(0,0)[b]{\smash{Dense\\ }}}%
    \put(0.56920649,0.54846418){\color[rgb]{0,0,0}\makebox(0,0)[b]{\smash{Dense}}}%
    \put(0.18700145,0.44697193){\color[rgb]{0,0,0}\makebox(0,0)[lb]{\smash{Dropout}}}%
    \put(0,0){\includegraphics[width=\unitlength,page=2]{dnn.pdf}}%
    \put(0.68938807,0.44798686){\color[rgb]{0,0,0}\makebox(0,0)[lb]{\smash{Speaker 1}}}%
    \put(0,0){\includegraphics[width=\unitlength,page=3]{dnn.pdf}}%
    \put(0.68938807,0.39521088){\color[rgb]{0,0,0}\makebox(0,0)[lb]{\smash{Speaker 2}}}%
    \put(0,0){\includegraphics[width=\unitlength,page=4]{dnn.pdf}}%
    \put(0.68938807,0.07957){\color[rgb]{0,0,0}\makebox(0,0)[lb]{\smash{Speaker C}}}%
    \put(0,0){\includegraphics[width=\unitlength,page=5]{dnn.pdf}}%
    \put(0.68938807,0.01664482){\color[rgb]{0,0,0}\makebox(0,0)[lb]{\smash{$\langle unk \rangle$}}}%
    \put(0,0){\includegraphics[width=\unitlength,page=6]{dnn.pdf}}%
    \put(0.08119379,0.51152099){\color[rgb]{0,0,0}\makebox(0,0)[lb]{\smash{LeakyReLU}}}%
    \put(0.29026782,0.51152099){\color[rgb]{0,0,0}\makebox(0,0)[lb]{\smash{LeakyReLU}}}%
    \put(0.5094911,0.51152099){\color[rgb]{0,0,0}\makebox(0,0)[lb]{\smash{Softmax}}}%
  \end{picture}%
\endgroup%

%% file: odyssey2018_article_arxiv.bbl
\begin{thebibliography}{10}

\bibitem{dehak2011front}
Najim Dehak, Patrick~J Kenny, R{\'e}da Dehak, Pierre Dumouchel, and Pierre
  Ouellet,
\newblock ``Front-end factor analysis for speaker verification,''
\newblock {\em IEEE Transactions on Audio, Speech, and Language Processing},
  vol. 19, no. 4, pp. 788--798, 2011.

\bibitem{mann2007}
Gideon~S Mann and Andrew McCallum,
\newblock ``Simple, robust, scalable semi-supervised learning via expectation
  regularization,''
\newblock in {\em ICML}, 2007, pp. 593--600.

\bibitem{nagrani2017}
Arsha Nagrani, Joon~Son Chung, and Andrew Zisserman,
\newblock ``{VoxCeleb}: A large-scale speaker identification dataset,''
\newblock {\em Interspeech}, pp. 2616--2620, 2017.

\bibitem{mauclair2006}
Julie Mauclair, Sylvain Meignier, and Yannick Esteve,
\newblock ``Speaker diarization: about whom the speaker is talking?,''
\newblock in {\em Odyssey: The Speaker and Language Recognition Workshop},
  2006, pp. 1--6.

\bibitem{elkhoury2012}
Elie El~Khoury, Antoine Laurent, Sylvain Meignier, and Simon Petitrenaud,
\newblock ``Combining transcription-based and acoustic-based speaker
  identifications for broadcast news,''
\newblock in {\em ICASSP}, 2012, pp. 4377--4380.

\bibitem{jousse2009}
Vincent Jousse, Simon Petit-Renaud, Sylvain Meignier, Yannick Esteve, and
  Christine Jacquin,
\newblock ``Automatic named identification of speakers using diarization and
  {ASR} systems,''
\newblock in {\em ICASSP}, 2009, pp. 4557--4560.

\bibitem{satoh1999}
Shin'ichi Satoh, Yuichi Nakamura, and Takeo Kanade,
\newblock ``Name-it: Naming and detecting faces in news videos,''
\newblock {\em Multimedia}, vol. 6, no. 1, pp. 22--35, 1999.

\bibitem{poignant2015}
Johann Poignant, Laurent Besacier, and Georges Qu{\'e}not,
\newblock ``Unsupervised speaker identification in {TV} broadcast based on
  written names,''
\newblock {\em IEEE/ACM Transactions on Audio, Speech, and Language
  Processing}, vol. 23, no. 1, pp. 57--68, 2015.

\bibitem{parkhi2015}
Omkar~M Parkhi, Andrea Vedaldi, and Andrew Zisserman,
\newblock ``Deep face recognition,''
\newblock in {\em British Machine Vision Conference}, 2015.

\bibitem{zhou2007multi}
Zhi-Hua Zhou and Min-Ling Zhang,
\newblock ``Multi-instance multi-label learning with application to scene
  classification,''
\newblock in {\em Advances in Neural Information Processing Systems}, 2007, pp.
  1609--1616.

\bibitem{zhou2012multi}
Zhi-Hua Zhou, Min-Ling Zhang, Sheng-Jun Huang, and Yu-Feng Li,
\newblock ``Multi-instance multi-label learning,''
\newblock {\em Artificial Intelligence}, vol. 176, no. 1, pp. 2291--2320, 2012.

\bibitem{zhang2014review}
Min-Ling Zhang and Zhi-Hua Zhou,
\newblock ``A review on multi-label learning algorithms,''
\newblock {\em IEEE Transactions on Knowledge and Data Engineering}, vol. 26,
  no. 8, pp. 1819--1837, 2014.

\bibitem{huang2014fast}
Sheng-Jun Huang, Wei Gao, and Zhi-Hua Zhou,
\newblock ``Fast multi-instance multi-label learning,''
\newblock in {\em 28th AAAI Conference on Artificial Intelligence}, 2014.

\bibitem{briggs2012rank}
Forrest Briggs, Xiaoli~Z Fern, and Raviv Raich,
\newblock ``Rank-loss support instance machines for {MIML} instance
  annotation,''
\newblock in {\em KDD}. ACM, 2012, pp. 534--542.

\bibitem{alumae2014recent}
Tanel Alum{\"a}e,
\newblock ``Recent improvements in {Estonian LVCSR},''
\newblock in {\em SLTU}, 2014.

\bibitem{meignier2010lium}
Sylvain Meignier and Teva Merlin,
\newblock ``{LIUM SpkDiarization}: an open source toolkit for diarization,''
\newblock in {\em CMU SPUD Workshop}, 2010.

\bibitem{chen1998}
Scott Chen and Ponani Gopalakrishnan,
\newblock ``Speaker, environment and channel change detection and clustering
  via the {Bayesian} information criterion,''
\newblock in {\em DARPA Broadcast News Transcription and Understanding
  Workshop}, 1998.

\bibitem{reynolds1998blind}
Douglas~A Reynolds, Elliot Singer, Beth~A Carlson, Gerald~C O'Leary, Jack~J
  McLaughlin, and Marc~A Zissman,
\newblock ``Blind clustering of speech utterances based on speaker and language
  characteristics,''
\newblock in {\em SLP}, 1998.

\bibitem{kaldi}
Daniel Povey, Arnab Ghoshal, Gilles Boulianne, Lukas Burget, Ondrej Glembek,
  Nagendra Goel, Mirko Hannemann, Petr Motlicek, Yanmin Qian, Petr Schwarz, Jan
  Silovsky, Georg Stemmer, and Karel Vesely,
\newblock ``The {Kaldi} speech recognition toolkit,''
\newblock in {\em ASRU}, 2011.

\bibitem{pyannote.metrics}
Herv\'e Bredin,
\newblock ``{pyannote.metrics: a toolkit for reproducible evaluation,
  diagnostic, and error analysis of speaker diarization systems},''
\newblock in {\em Interspeech}, 2017.

\bibitem{tsab}
Tanel Alum{\"a}e and Ahti Kitsik,
\newblock ``{TSAB} -- web interface for transcribed speech collections,''
\newblock in {\em Interspeech}, 2011.

\end{thebibliography}
